\newcommand{\kms}{{~\rm km\; s^{-1}}}
\newcommand{\cc}{{~\rm cm^{-3}}}

\newcommand{\s}{{~\rm s}}
\newcommand{\km}{{~\rm km}}
\newcommand{\g}{{~\rm g}}
\newcommand{\K}{{~\rm K}}
\newcommand{\erg}{{~\rm erg}}
\newcommand{\yr}{{~\rm yr}}
\newcommand{\Myr}{{~\rm Myr}}

\newcommand{\kpc}{{~\rm kpc}}

\documentclass[12pt,preprint]{aastex}
\shorttitle{Fat bubbles}
\shortauthors{Sternberg \& Soker}

\begin{document}

\title{INFLATING FAT BUBBLES IN CLUSTERS OF GALAXIES BY PRECESSING MASSIVE
SLOW JETS}

\author{Assaf Sternberg\altaffilmark{1},  Noam Soker\altaffilmark{1}}

\altaffiltext{1}{Department of Physics,
Technion$-$Israel Institute of Technology, Haifa 32000, Israel;
phassaf@techunix.technion.ac.il; soker@physics.technion.ac.il}

\begin{abstract}
We conduct hydrodynamical numerical simulations and find that precessing
massive slow jets can inflate fat bubbles, i.e., more or less spherical bubbles,
that are attached to the center of clusters of galaxies. To inflate a fat bubble the
jet should precess fast. The precessing angle $\theta$ should be large, or
change over a large range $ 0 \le \theta \le \theta_{\max} \sim 30-70 ^\circ$
(depending also on other parameters), where $\theta=0$ is the
symmetry axis. The constraints on the velocity and mass outflow rate are
similar to those on wide jets to inflate fat bubbles. The velocity should be
$v_j \sim 10^4 \kms$,  and the mass loss rate of the two jets should be
$ 2 \dot M_j \simeq 1-50  \dot M_\odot \yr^{-1} $. These results, and our
results from a previous paper dealing with slow wide jets, support the claim
that a large fraction of the feedback heating in cooling flow clusters and
in the processes  of galaxy formation is done by slow massive jets.
\end{abstract}


\section{INTRODUCTION}
\label{sec:intro}

Many of the X-ray deficient bubbles in galaxies and clusters of galaxies
reside very close to the center of the cluster (or galaxy) and are fully
or partially surrounded by a dense shell, e.g., Perseus (Fabian et al. 2000),
Abell~2052, (Blanton et al. 2003), Abell~4059 (Heinz et al. 2002), and HCG~62
(Vrtilek et al. 2002; Morita et al. 2006). We term these more or less
spherical bubbles `fat bubbles'. {{{ Fat bubbles are defined by the 
following properties: (1) Fat bubbles come in pairs, each on 
opposite sides of the equatorial plane. In some cases there is departure from 
axisymmetry, and they are not exactly opposite. (2) They touch each other 
at the center, and by that form an hourglass structure (like the figure '8'). 
One bubble of the hourglass structure is referred to as a fat bubble. 
(3) The density inside the bubbles is much lower than that of their 
surroundings (ambient gas). (4) They are fully or partially surrounded by 
relatively thin shell that is denser than the surroundings. (5) In some cases 
their boundary, on the far side from the center, is open. In 
classifying planetary nebulae, for example, this structure is refereed to as 
a bipolar nebula. There are many such planetary nebulae that are well resolved 
in the visible band, and the bipolar structure is well defined. In many cases 
there are similar structures in planetary nebulae and in clusters of galaxies 
(Soker \& Bisker 2005).}}} The best examples of the hourglass type of 
structure we aim to study are the bubbles in Perseus (Fabian et al. 2000) and 
in A~2052, (Blanton et al. 2001). {{{Other cases are cited above and in 
Soker \& Bisker 2006. If the bubble rises through the ICM and do not touch 
each other any mode, then they are not 'fat' any more because they don't form 
an hourglass structure anymore. We estimate that $\sim 30-50 \%$ of the 
bubbles are, or were, fat bubbles.}}}

In recent years, two and three-dimensional hydrodynamical simulations of jets
and bubbles in clusters of galaxies, were conducted to study different
aspects of their interaction with the intra-cluster medium (ICM), such as heating 
the ICM (e.g., Basson \& Alexander, 2003; Heinz \& Churazov, 2005; Reynolds et al.,
2005; Heinz et al.,
2006; Vernaleo \& Reynolds, 2006; Binney et al., 2007; Ruszkowski et al.,
2007; Alouani Bibi et al., 2007; Br\"uggen et al., 2007). In our studies we
aim to understand the conditions leading to the formation of fat bubbles.

In previous papers (Soker 2004, 2006; Sternberg et al. 2007, hereafter Paper I)
we proposed that in order to inflate fat bubbles, either the jet's opening
angle has to be large, i.e., wide jets, or the jet is narrow but its axis has to 
change its direction. The change in direction can result from precession (Soker 2004,
2006), random change (Heinz et al. 2006), or a relative motion  between the
ICM and the active galactic nucleus (AGN; Loken et al. 1995; Soker \& Bisker
2006; Rodr\'iguez-Mart\'inez et al. 2006).

In Paper I we conducted
two-dimensional hydrodynamical simulations of wide jets expanding in the
ICM. We found that wide jets can indeed inflate fat
bubbles that reside very close to the center of the cluster. For this to
occur, we found that the jets should have high momentum flux. Typically, the
half opening angle should be $\alpha \ga 50^\circ$, and the large momentum
flux requires a jet speed of $v_j \sim 10^4 \km \s^{-1}$, i.e., highly
non-relativistic, but supersonic ($v_j \simeq 3000-3 \times 10^4 \km \s^{-1}$).
Narrow relativistic jets can exist in parallel with the slow wide outflow, but
they will not lead to the inflation of fat bubbles.
The inflation process involves large vortices and
local instabilities which mix some ICM  with the hot bubble. These results
predict that most of the gas inside the bubble has a temperature of
$3 \times 10^8 \la T_{b} \la 3 \times 10^{9} \K$, and that large quantities
of the cooling gas in cooling flow clusters are expelled back to the
intra-cluster medium and heated up (Soker \& Pizzolato 2005). In Paper I we
suggested that the magnetic fields and relativistic electrons that produce the
synchrotron radio emission might be formed in the shock wave of the
non-relativistic jet. Motivated by our earlier results, in this paper we
examine the inflation of fat bubbles by narrow precessing jets.

\section{NUMERICAL METHOD AND SETUP}
\label{sec:numerics}

The simulations were performed using the \emph{Virginia} \emph{Hydrodynamics-I}
code (VH-1; Blondin et al. 1990; Stevens et al. 1992), as described in Paper I.
We simulated a three-dimensional axisymmetric flow, so practically we simulated
a quarter of the meridional plane using the two-dimensional version of the
code in spherical coordinates.
The symmetry axis of all plots shown in this paper is along the horizontal axis (the x
axis) while the equitorial plane is vertical.
Radiative cooling and gravity were not included, since the total time of the
simulation, $t_{sim}\sim 10^7 \yr$,  is somewhat shorter than the
gravitational time scale, and much shorter than the radiative cooling time.
This preliminary report aims to emphasize the jet properties that
determine whether or not the required bubble is inflated, hence, these omissions
are justified. 

\begin{figure}
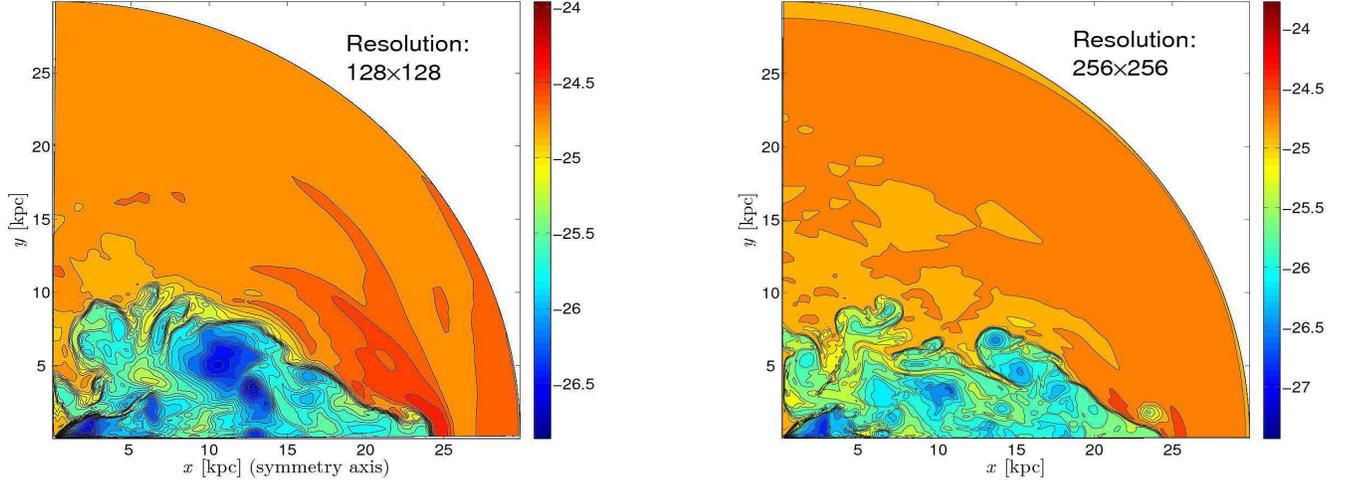
  
\hskip -2.0 cm  
{\includegraphics[scale=0.29]{assafp2f1a.eps2}}  
\hskip -1.5 cm  
{\includegraphics[scale=0.29]{assafp2f1b.eps2}}  
\caption{Density maps for two runs with the same parameters but different 
resolution ($128\times128$ and $256\times256$) shown at the same time 
($t=27\Myr$). The small scale behavior might differ, but the large scale 
behavior, which is what we are interested in, does not differ. We emphasize that 
the x axis is the symmetry axis. In all figures the density is given in $\rm{g}\;\rm{cm}^{-3}$ and in log scale.}
\label{resolution}
\end{figure}

{{{We used the $\beta$ model (with $\beta=1/2$) as the initial density profile of the ICM,
\begin{equation}
\rho_{\rm ICM}= \rho_c [(1+(r/r_0)^2]^{-3/4},
\label{rho}
\end{equation}
with $\rho_c = 2.16 \times 10^{-25} \g\cc$ and $r_0=100 \kpc$ (see Paper I and 
references therein)}}}. The ICM temperature is $2.7 \times 10^7 \K$. The box size 
used in our simulations was $30 \kpc \times 30 \kpc$ (one quarter of the meridional 
plane).{{{ We used a $128\times128$ evenly spaced grid. As discussed in Paper I, higher 
resolution does not change the large scale behavior, thus, this resolution is 
sufficient for our study. In Figure \ref{resolution} we present density maps
(logarithem of the density. The density is given in $\rm{g}\;\rm{cm}^{-3}$.) of two runs with the same parameters but different resolution ($128\times128$ and 
$256\times256$). This figure clearly shows that the large scale behavior does 
not change significantly with the increase in resolution.}}}  Moreover, we note that 
in the zone relevant to our simulations, $r \la 30 \kpc$, the assumption of a constant temperature is reasonable (see fig 1. of Pizzolato \& Soker 2005).

The narrow jet was injected at a radius of $0.1 \kpc$, with constant mass
flux $\dot M_j$ (per one jet) and a constant radial velocity $v_j$,
inside a half opening angle $\alpha = 5^\circ$. Therefore,
the total kinetic power of one jet is $\dot E_j=\dot M_j v_j^2/2$. . The
symmetry axis of the jet is at an angle $\theta(t)$ in respect to the
symmetry axis of the  problem (the x axis). This is the precessing angle
which is a function of time. Because of the axisymmetric nature of our
problem, the meaning of a precessing jet in these simulations is that the
narrow jet precess around the symmetry axis very rapidly. Namely, the
precessing period around the symmetry axis is much shorter than any other
relevant time scale in the problem, e.g., the time scale over which $\theta$
is changing.

Due to the numerical nature of the jets injection, in some cases
a fraction of the injected mass does not succeed in mounting our grid and
remains in the first cells of the grid. In the next time step the jet
properties are reinserted into these cells and this mass is lost. Moreover,
if the jet is precessing, then the movement of the injection zone my
increase/decrease the effective opening angle of the jet, in effect changing 
$\dot M_j$, such that only after the simulation we can determine the mass loss 
rate and kinetic luminosity of the jet. The kinetic luminosity of the jet is constant, and is equal to $5PV/2t$, where $P$ is the pressure and $V$ is the volume
inflated after time $t$. We calculated this value at several times for each case, and was taken to be the kinetic luminosity. All {{{tables and}}} plot captions specify the estimated jet luminosity.
We consider three basic types of precessing jets:
\begin{description}
\item[(i)] Fixed precessing angle, i.e., $\theta$ is constant.
\item[(ii)] A constant rate of change in the precession angle, i.e., at 
constant time interval $d \theta/dt$ is constant for $\theta<\theta_{\rm max}$.
\item[(iii)] Random precession, i.e., the jet axis has the same probability to
take any direction within a maximum angle $\theta_{\rm max}$. This is done by
taking  $d(\cos\theta)/dt$ to be constant.
\end{description}
{{{ We studied 12 cases per each precession type (fixed, constant rate 
change, and random). The parameters of the different cases are given in 
Tables \ref{tab_fix}-\ref{tab_rand}.}}}

We used slow massive jets, as have been used before in a number of numerical
studies (e.g., Paper I; Alouani Bibi et al. 2007).
We further discuss the usage of slow massive jets in section 6.

\begin{table}
\hskip -1.5 cm
\begin{tabular}{|c|c|c|c|c|c|}
\hline
\multicolumn{6}{|c|}{Precession at a fixed angle} \\
\hline
\hline
Run & $\theta (\rm{deg})$ & $v_j (\frac{\rm{Km}}{\rm{s}})$ & 
$L_j (10^{44}\,\frac{\rm{erg}}{\rm{s}})$ & Morphology & Figure \\
\hline
\hline
Fix1 & $15^\circ$ & 7750 & 0.24 & Thin and narrow jet and 
cocoon & - \\
Fix2 & $15^\circ$ & 7750 & 1.1 & Thin and narrow jet and 
extensive cocoon & Fig \ref{const_15deg_E^44}\\
Fix3 & $15^\circ$ & 23250 & 0.15 & Very Thin and narrow jet and 
cocoon & - \\
Fix4 & $15^\circ$ & 23250 & 1.8 & Thin and narrow jet and 
extensive cocoon & Fig \ref{const_not_good}\\
Fix5 & $30^\circ$ & 7750 & 0.14 & Fat Bubble & Fig \ref{const_30deg_E^43} \\
Fix6 & $30^\circ$ & 7750 & 1.8 & Fat bubble & Fig \ref{const_30deg_E^44}\\
Fix7 & $30^\circ$ & 23250 & 1.2 & First torus then spherical cavity 
at center of cluster & Fig \ref{const_not_good}\\
Fix8 & $30^\circ$ & 69750 & 1.2 & First torus then an ellipsoid cavity 
at center of cluster & - \\
Fix9 & $60^\circ$ & 7750 & 0.2 & First torus then an ellipsoid cavity 
at center of cluster & - \\
Fix10 & $60^\circ$ & 7750 & 1.8 & Both tori merge to create a 'doughnut' & 
Fig \ref{const_not_good} \\
Fix11 & $60^\circ$ & 23250 & 0.14 & First torus then an ellipsoid cavity 
at center of cluster & - \\
Fix12 & $60^\circ$ & 23250 & 2.2 & First torus then spherical cavity 
at center of cluster & - \\
\hline
\end{tabular}
\caption{Parameters of the fixed angle precession runs, where, $\theta$ is 
the angle (in degrees) between the symmetry axis of the jet and the symmetry 
axis of the problem, $v_j$ is the jet velocity in $\rm{Km}\;\rm{s}^{-1}$, 
$L_j$ is the jet luminosity in $\rm{erg}\;\rm{s}^{-1}$ (per one jet). The 
cluster sound speed is $775\,\rm{Km}\;\rm{s}^{-1}$.}
\label{tab_fix} 
\end{table}

\begin{table}
\hskip -1.2 cm
\begin{tabular}{|c|c|c|c|c|c|}
\hline
\multicolumn{6}{|c|}{Constant rate of change in the precession angle} \\
\hline
\hline
Run & $T (\rm{Myr})$ & $v_j (\frac{\rm{Km}}{\rm{s}})$ & 
$L_j (10^{44}\,\frac{\rm{erg}}{\rm{s}})$ & Morphology & Figure \\
\hline
\hline
Con1 & 0.1 & 7750 & 1.9 & Fat bubble & Fig \ref{uniform_T0_1_E^44} \\
Con2 & 0.1 & 23250 & 1.1 & Thin jet shedding large vortecies & 
Fig \ref{uniform_T0_1_E^44_fast}\\
Con3 & 0.1 & 69750 & 1.8 & Very thin jet and extensive cocoon & - \\
Con4 & 1 & 7750 & 2 & Narrow and elongated cavity (not a fat bubble!!!) 
& - \\
Con5 & 1 & 23250 & 1.5 & Narrow and elongated cavity &
- \\
Con6 & 1 & 69750 & 1.2 & Narrow and elongated cavity & - \\
Con7 & 5 & 7750 & 1.6 & Narrow and elongated cavity extensive backflow & 
Fig \ref{uniform_T5_E^44}\\
Con8 & 5 & 23250 & 1.6 & Narrow jet with extensive backflow & - \\
Con9 & 5 & 69750 & 2 & Narrow jet with extensive backflow & - \\
Con10 & 30 & 7750 & 1.8 & Elongated clumpy cavity, extensive backflow  & 
- \\
Con11 & 30 & 23250 & 1.5 & Narrow jet with extensive clumpy cocoon & - \\
Con12 & 30 & 69750 & 1.6  & Narrow jet with narrow clumpy cocoon & - \\
\hline
\end{tabular}
\caption{Parameters of the constant rate of change in precession angle runs. 
Same parameters as in Table \ref{tab_fix}. In addition, $T$ is the precession 
period in Myr.}
\label{tab_unif} 
\end{table}

\begin{table}
\hskip -1.2 cm
\begin{tabular}{|c|c|c|c|c|c|}
\hline
\multicolumn{6}{|c|}{Random change in the precession angle} \\
\hline
\hline
Run & $T (\rm{Myr})$ & $v_j (\frac{\rm{Km}}{\rm{s}})$ & 
$L_j (10^{44}\,\frac{\rm{erg}}{\rm{s}})$ & Morphology & Figure \\
\hline
\hline
Ran1 & 0.1 & 7750 & 1.4 & Fat bubble & Fig \ref{rand_T0_1} \\
Ran2 & 0.1 & 23250 & 2 & Thin jet shedding very large vortecies & 
- \\
Ran3 & 0.1 & 69750 & 1.5 & Thin jet and cocoon, extensive backflow & - \\
Ran4 & 1 & 7750 & 1.8 & Elongated cavity & - \\
Ran5 & 1 & 23250 & 1.6 & Elongated cavity &
- \\
Ran6 & 1 & 69750 & 1.5 & Narrow and elongated cavity extensive backflow & - \\
Ran7 & 5 & 7750 & 1.5 & Elongated clumpy cavity & 
- \\
Ran8 & 5 & 23250 & 1.8 & Elongated clumpy cavity & - \\
Ran9 & 5 & 69750 & 2 & Elongated clumpy cavity, extensive backflow & - \\
Ran10 & 30 & 7750 & 1.8 & First torus then fat bubble  & Fig \ref{rand_T30} \\
Ran11 & 30 & 23250 & 1.5 & First torus then elongate cavity, extensive 
backflow & - \\
Ran12 & 30 & 69750 & 1.6  & Torus then narrow jet with clumpy cocoon & - \\
\hline
\end{tabular}
\caption{Parameters of the random change in precession angle runs. Same as in 
Table \ref{tab_unif}.}
\label{tab_rand} 
\end{table}

\section{RESULTS: PRECESSION AT A CONSTANT ANGLE}
\label{sec:results1}

\begin{figure}
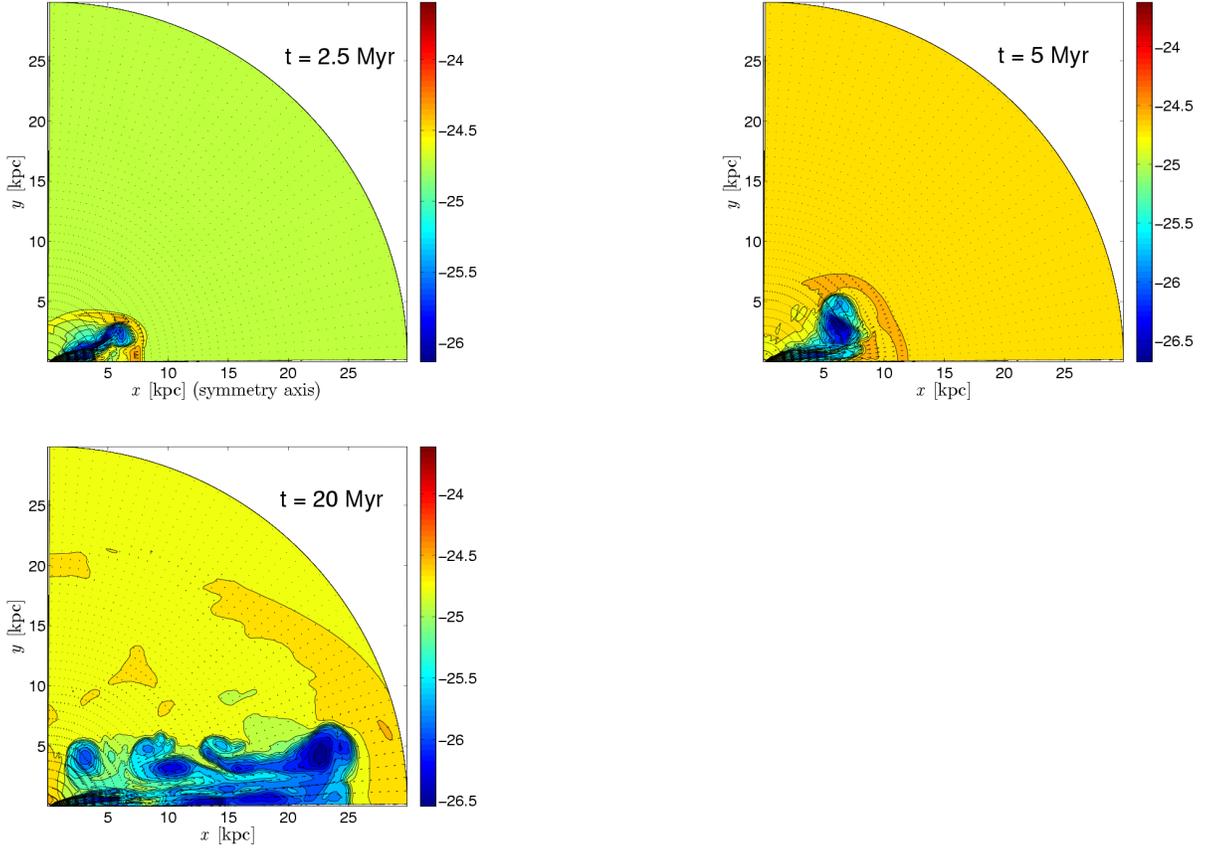
  
{\includegraphics[scale=0.29]{assafp2f2a.eps2}}  
{\includegraphics[scale=0.29]{assafp2f2b.eps2}}  
{\includegraphics[scale=0.29]{assafp2f2c.eps2}}
\caption{Density maps (logarithem of the density [$\rm{g}\;\rm{cm}^{-3}$]) for a 
jet at a fixed precession angle $\theta=15^\circ$ (Fix2, see Table \ref{tab_fix}), 
given at three different 
times ($t=2.5,\:5,$ and $20\Myr$). The jet has a half opening angle of
$\alpha=5^\circ$, an injected velocity of $v_j=7750 \km\s^{-1}$, and power of
(one jet) $\dot E_j\simeq 1.1\times 10^{44} \erg\s^{-1}$. Only one quarter
of the meridional plane is showed, as the other three are symmetric to it. The
$x$-axis (horizontal) is the symmetry axis, while the $y$-axis (vertical) is
in the equatorial plane. The arrows represent the velocity of the flow:
$0.1c_s<v_j \leq 0.5c_s$ (shortest), $0.5c_s<v_j \leq c_s$,
$c_s<v_j \leq 5c_s$, and $5c_s<v_j \leq 10c_s$ (longest in this case).}
\label{const_15deg_E^44}
\end{figure}

In Figure~\ref{const_15deg_E^44} we show the density map (logarithem of the density. 
The density is given in $\rm{g}\;\rm{cm}^{-3}$.) 
at different times
for a jet with a fixed precession angle of $\theta=15^\circ$ in respect
to the symmetry axis (Fix2, see Table \ref{tab_fix}). The jet has a half 
opening angle of $\alpha=5^\circ$, $v_j=7750 \km\s^{-1}$, and
$\dot E_j\simeq 1.1\times 10^{44} \erg\s^{-1}$. For this case the mass
injection rate into one jet is $\dot M_j\simeq 6M_\odot
\yr^{-1}$. Namely, the two jets expel mass back to the ICM at a high rate of
$\sim 12 M_\odot \yr^{-1}$. At first the jet inflates a cavity of low
density matter in the shape of a torus. At later times the jet is bent
towards the symmetry axis, thereafter, it continues it's propagation in a
manner quit similar to that of a non precessing jet with a $\sim 20^\circ$
half opening angle (see Figure 1 model 2 in Paper I). In this case, it is
obvious to see that a fat bubble was not inflated. The arrows in the plot
represent the velocity of the flow. For visual clarity we divided the
velocities into groups, each represented by an arrow of a predetermined
length:
\begin{description}
\item[(i)] $0.1c_s<v_j \leq 0.5c_s$ - shortest,
\item[(ii)] $0.5c_s<v_j \leq c_s$,
\item[(iii)] $c_s<v_j \leq 5c_s$,
\item[(iv)] $5c_s<v_j \leq 10c_s$ - longest for the $M=10$ case,
\item[(v)] $10c_s<v_j \leq 30c_s$ - longest for the $M=30$ case,
\end{description}
where $c_s=775\kms$ is the speed of sound. For visual clarity we also omitted
velocities of $v_j\leq 0.1 c_s$. This division is true for all plots shown in
this paper.

\begin{figure}
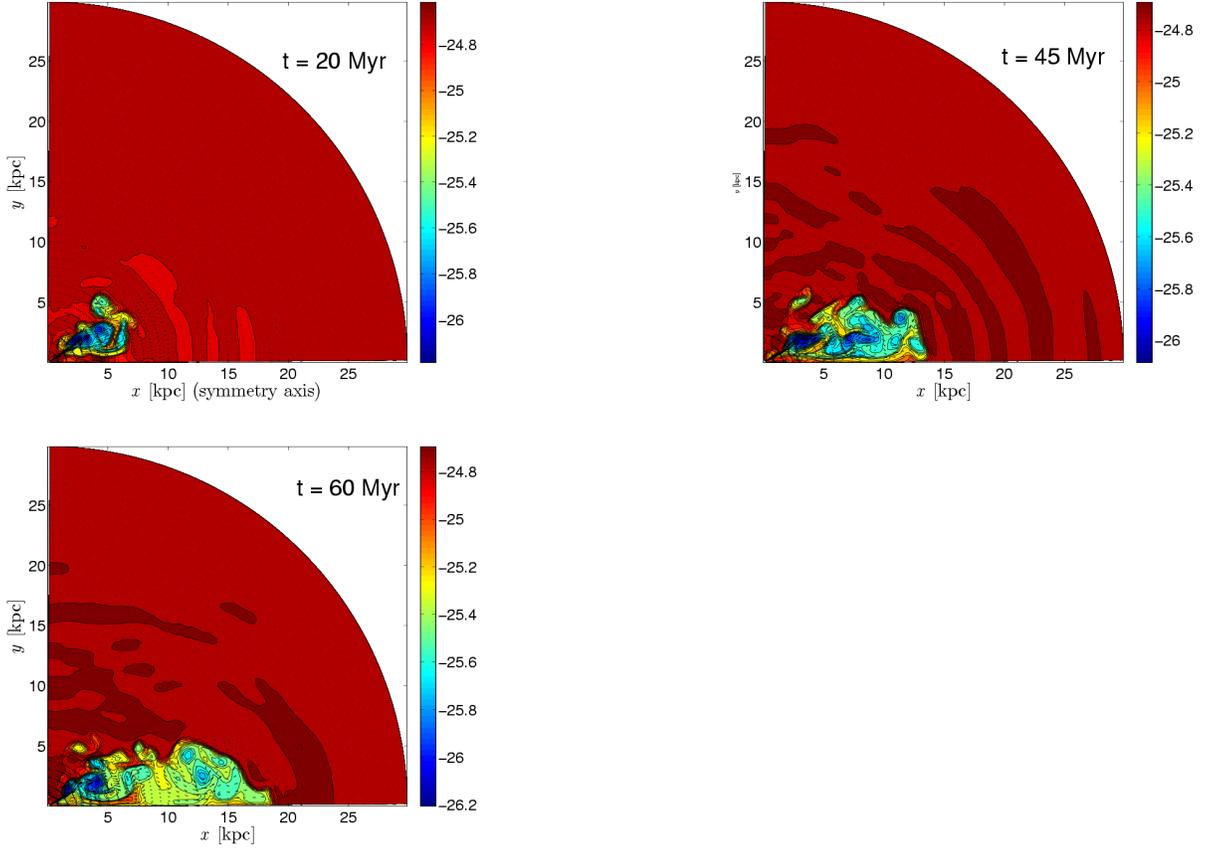
  
{\includegraphics[scale=0.29]{assafp2f3a.eps2}}
{\includegraphics[scale=0.29]{assafp2f3b.eps2}}
{\includegraphics[scale=0.29]{assafp2f3c.eps2}}
\caption{Density maps for a jet at a fixed precession
angle of $\theta=30^\circ$ (Fix5, see Table \ref{tab_fix}), with a half 
opening angle of $\alpha=5^\circ$,
$v_j=7750 \km\s^{-1}$, and $\dot E_j\simeq 1.4\times 10^{43} \erg\s^{-1}$.
The cavity inflated by the jet reaches a more or less spherical shape at
$t \sim 45 \Myr$, and remains spherical, though a little elongated even at
$t=60 \Myr$. The equatorial plane (vertical in the figures) remains devoid of 
low density matter, in good agreement with observations. There are no numerical
fluctuations in the equatorial boundary either, unlike along the symmetry axis
(horizontal axis) The arrows represent the velocity of the flow as in Figure 
\ref{const_15deg_E^44}. }
\label{const_30deg_E^43}
\end{figure}

\begin{figure}  
{\includegraphics[scale=0.29]{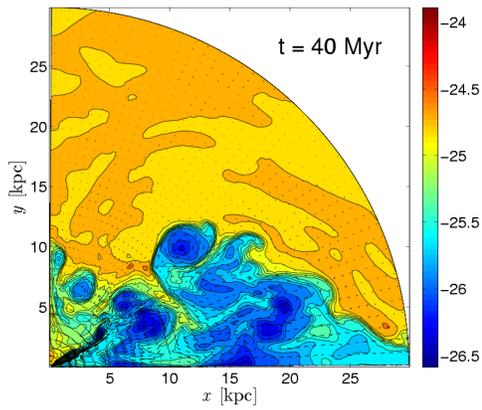}}
\caption{Density maps for a jet at a fixed precession
angle of $\theta=30^\circ$, with a half opening angle of $\alpha=5^\circ$,
$v_j=7750 \km\s^{-1}$ (Fix6, see Table \ref{tab_fix}), 
and $\dot E_j\simeq 2\times 10^{44} \erg\s^{-1}$. The
jet inflates a more or less spherical cavity. The bubble is shown at
$t=40 \Myr$. There is a non-negligible flow of low density matter towards the
equatorial plane, which is not in good agreement with observations. The arrows
represent the velocity of the flow as in Figure \ref{const_15deg_E^44}..}
\label{const_30deg_E^44}
\end{figure}

In Figure~\ref{const_30deg_E^43} we show the density map at different times
for a jet at a fixed precession angle of $\theta=30^\circ$ (Fix5, see 
Table \ref{tab_fix}), with a half opening angle of $\alpha=5^\circ$, 
$v_j=7750 \km\s^{-1}$, and $\dot E_j\simeq 1.4\times 10^{43} \erg\s^{-1}$. 
In a similar manner to the
former case, the jet initially inflates a torus shaped cavity, and later it
is bent towards the symmetry axis. In contrast to the former case we see that at 
$t = 45$ Myr the jet has inflated a bubble that together with the contra-bubble will form a bipolar structure. The bubble in this case is much closer to being spherical 
than bubbles in runs that we consider unsuccessful in forming fat bubbles (they don't
form a bipolar structure). Even at $t=60 \Myr$ the cavity can still be termed an elongated bubble. In
contrast to the bubbles we showed in Paper I, there is no
flow of low density matter towards the equatorial plane, which is in good
agreement with observations. For the same parameters, but with a luminosity of
$\sim 2\times 10^{44} \erg\s^{-1}$ (Fix6, see Table \ref{tab_fix}), 
we also got a fat bubble of low
density, as shown in Figure~\ref{const_30deg_E^44}. In contrast to the lower
luminosity case, we see a non-negligible flow of low density matter towards
the equatorial plane, which is not in very good agreement with observations.
At early time a torus, rather than a fat bubble, is inflated. At present, we
know of no observations of such bubbles.

\begin{figure}
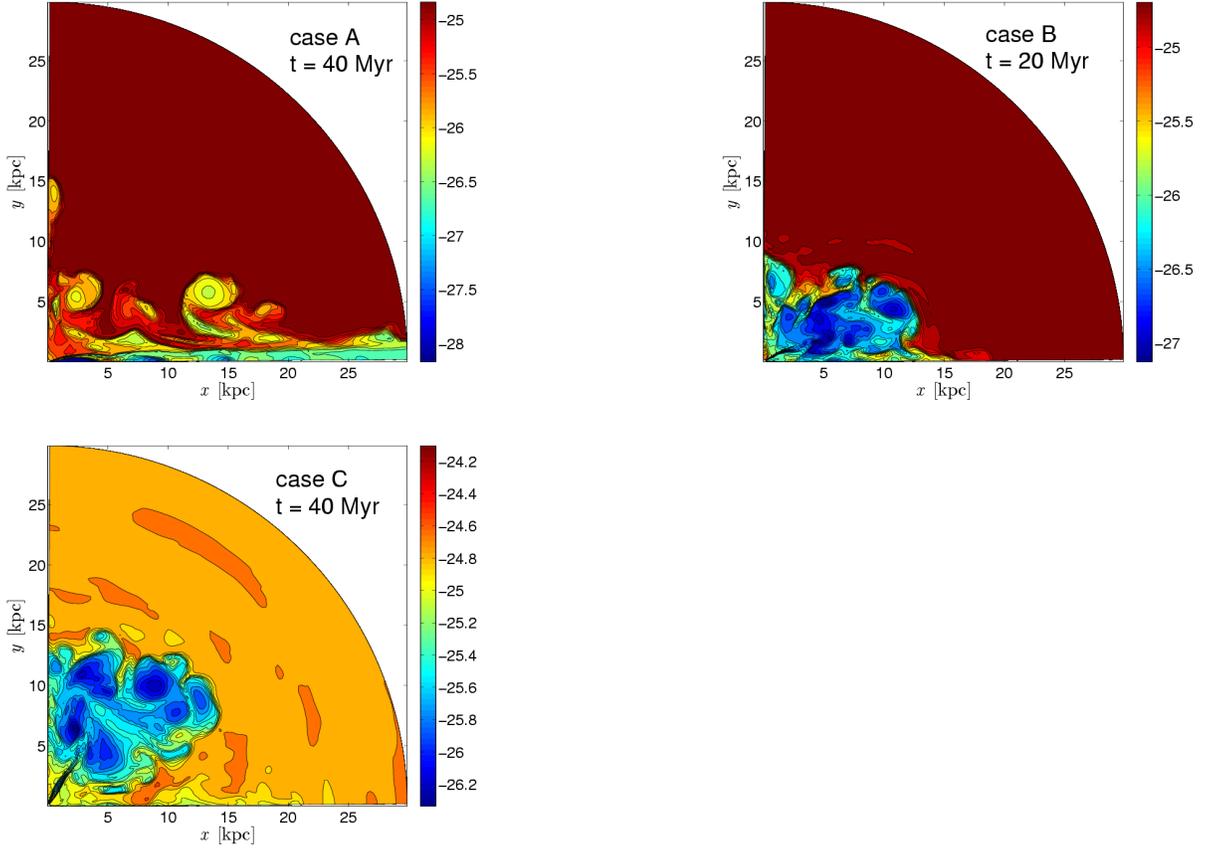
  
{\includegraphics[scale=0.29]{assafp2f5a.eps2}}
{\includegraphics[scale=0.29]{assafp2f5b.eps2}}
{\includegraphics[scale=0.29]{assafp2f5c.eps2}}
\caption{Density maps for three different cases of jets at a fixed
precession angles with a half opening angle of $\alpha=5^\circ$. In the upper
panel (case A: Fix4,see Table \ref{tab_fix}) we show a high velocity jet 
($v_j = 23250 \kms$) with a small precession angel ($\theta=15^\circ$), and 
jet luminosity of $\dot E_j\simeq 1.8\times 10^{44} \erg\s^{-1}$.
In the middle panel (case B: Fix7,see Table \ref{tab_fix}) we show high 
velocity jet ($v_j = 23250 \kms$) with an intermediate precession angel 
($\theta=30^\circ$), and jet luminosity of $\dot E_j\simeq 1.2\times 10^{44} 
\erg\s^{-1}$. In the lower panel (case C: Fix10,see Table \ref{tab_fix}) we 
show a low velocity jet ($v_j = 7750 \kms$) with a large precession angel 
($\theta=60^\circ$) and jet luminosity of 
$\dot E_j\simeq 1.8\times 10^{44} \erg\s^{-1}$. In all three cases a fat 
bubble was not inflated.}
\label{const_not_good}
\end{figure}

In most of the cases that we ran we ended up with either a narrow jet, as
shown in upper panel of Figure~\ref{const_not_good} (Fix4,see Table 
\ref{tab_fix}), or with a fat bubble with a significant flow of low 
density matter to the equatorial plane {{{(in effect a spherical or 
ellipsoid cavity at the center of the cluster)}}}, as shown in the middle 
panel of Figure~\ref{const_not_good} (Fix7, see Table \ref{tab_fix}). In some 
of the cases with $\theta=60^\circ$ (Fix10,see Table \ref{tab_fix}) the torus 
inflated by the simulated jet merged
at the equatorial plane with the torus of the unsimulated jet (we remind the
reader that we imposed reflecting boundary conditions in the equatorial plane).
In effect we got a large low density torus, with an elliptic cross section,
at the equator as shown in the lower panel of Figure~\ref{const_not_good}.
For the sake of clarity we state that all of these cases are not in agreement
with observations. Therefore, large constant precessing angle
$\theta \ga 40 ^\circ$, cannot form the observed fat bubbles. In all cases
the typical temperature of the low density gas in the cavities or cocoon was
$10^8 \K \la T_b \la 10^9 \K$.

\section{RESULTS: A CONSTANT RATE OF CHANGE IN THE PRECESSION ANGLE}
\label{sec:results2}

\begin{figure}
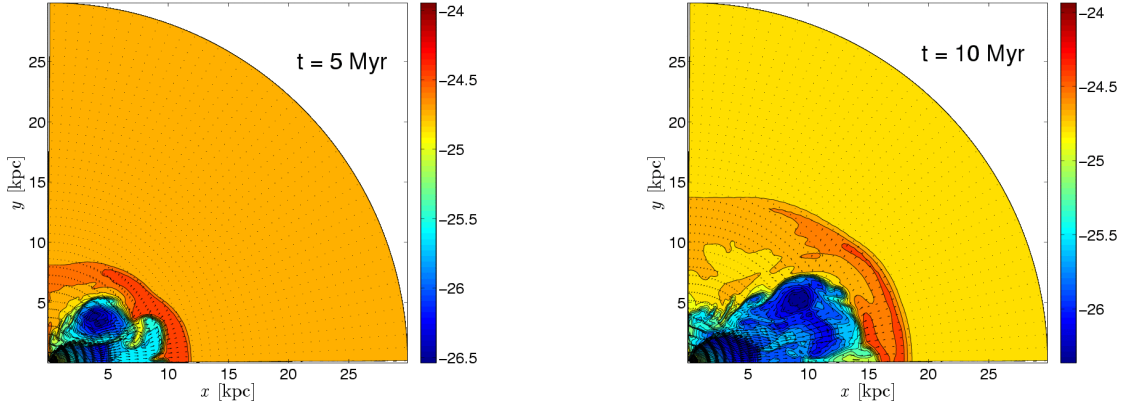
  
{\includegraphics[scale=0.29]{assafp2f6a.eps2}}
\hskip -1.0 cm  
{\includegraphics[scale=0.29]{assafp2f6b.eps2}}
\caption{Density maps for a jet with a uniformly changing precession angle,
$5^\circ\leq\theta\leq65^\circ$, with a precession period of
$T_{\rm prec}=0.1\Myr$, half opening angle of $\alpha=5^\circ$,
$v_j=7750 \km\s^{-1}$, and $\dot E_j\simeq 1.9\times 10^{44} \erg\s^{-1}$
(Con1,see Table \ref{tab_unif}). The arrows represent the velocity of the flow
as in Figure \ref{const_15deg_E^44}. }
\label{uniform_T0_1_E^44}
\end{figure}

\begin{figure}  
{\includegraphics[scale=0.29]{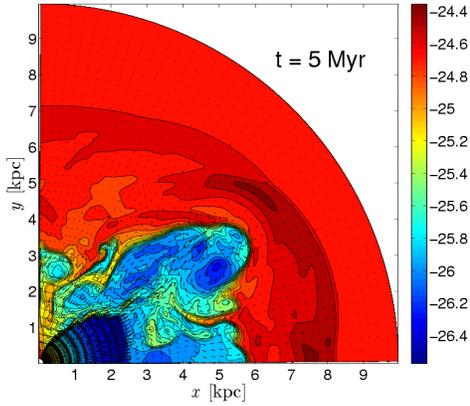}}
\caption{A case of a wide jet ($\alpha=70^\circ$)  instead of a precessing
jet, taken from Paper I. The parameters of this run are:
$v_j=7750 \km\s^{-1}$, and $\dot E_j\simeq 0.65\times 10^{44} \erg\s^{-1}$.
This plot shows the bubble  at $t=5\Myr$. The arrows represent the velocity
of the flow as in Figure \ref{const_15deg_E^44}.}
\label{wide}
\end{figure}

Figure \ref{uniform_T0_1_E^44} shows the density maps for a precessing jet
with a constant rate of change in the precession angle (i.e., 
$\theta\propto t$, where the symmetry axis of the jet changes in the range
$5^\circ \leq \theta \leq65^\circ$). We remind the reader that in all cases 
presented here, the jet is assumed to precesses rapidly around the symmetry 
axis, so it rotates in the $\phi$ direction many times while $\theta$ is 
being changed; the $\phi$ coordinate is not calculated in the simulations.
The case shown in this figure has a precession period $T_{\rm prec}=0.1 \Myr$
(Con1, see Table \ref{tab_unif}). This period is much shorter than the typical
expansion time of the bubble formed. As a result of that the narrow jet's
interaction with the ICM resembles that of a wide opening angle jet with a
half opening angle of $\sim 70^\circ$ (Paper I). For comparison, in figure
\ref{wide} we show the case of a wide angle jet taken from Paper I. As can
be seen in figure \ref{uniform_T0_1_E^44}, at short times of ($t \la 5 \Myr$)
the low density bubble is more or less spherical and there is almost no flow
of low density matter to the equatorial plane. At longer times of
$t\sim 10\Myr$, the flow of low density matter to the equatorial plane is
substantial, though the cavity itself can still be termed a fat bubble.
We conclude that for these parameters a fat bubble is formed.
The difference in the volume of the bubbles is due to the fact that the
actual $\dot{M}$ of the wide angle case was slightly smaller then that
of the precessing case, and therefore the bubble inflated by it was slightly
smaller (this is due to the numerics associated with the jet injection, as
elaborated in section \ref{sec:numerics}).

\begin{figure}
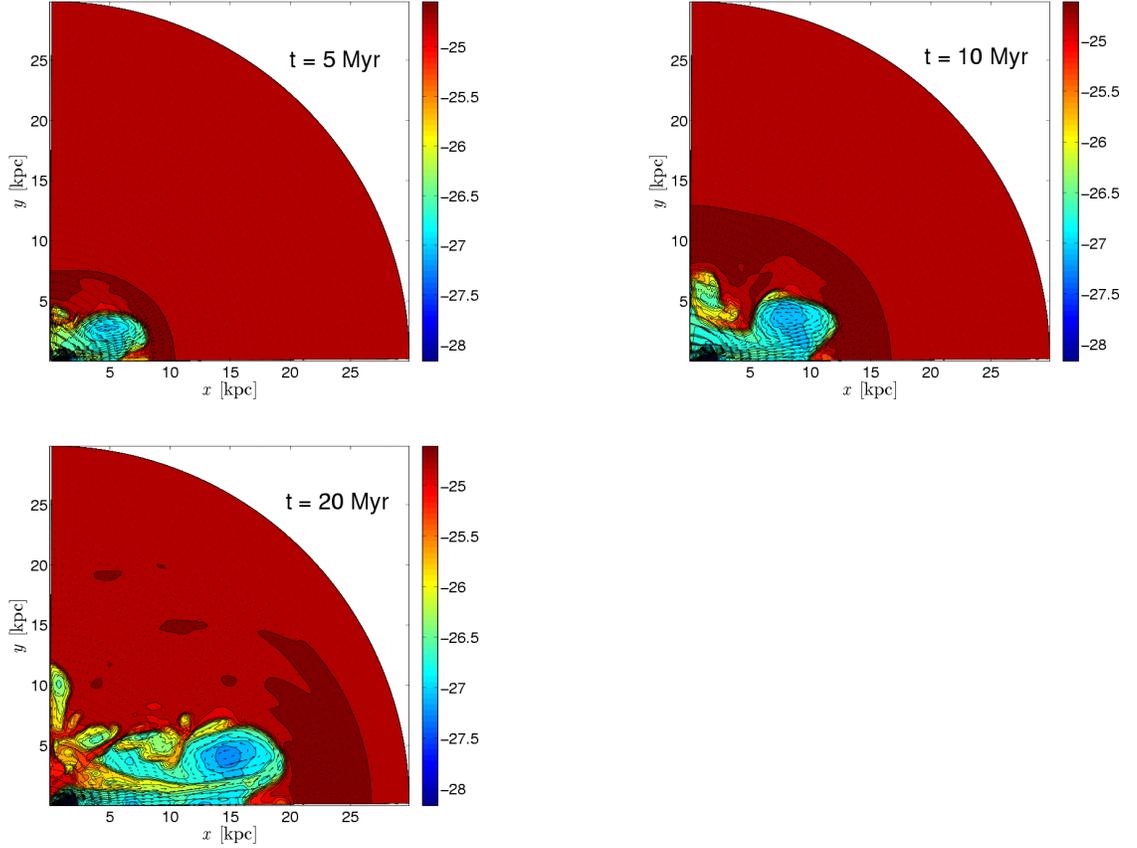
  
{\includegraphics[scale=0.29]{assafp2f8a.eps2}}
\hskip -1.0 cm  
{\includegraphics[scale=0.29]{assafp2f8b.eps2}}
\hskip -1.0 cm  
{\includegraphics[scale=0.29]{assafp2f8c.eps2}}
\caption{Like figure \ref{uniform_T0_1_E^44} but for a faster jet with
$v_j=23250 \km\s^{-1}$, i.e., $5^\circ\leq\theta\leq65^\circ$,
$T_{\rm prec}=0.1\Myr$, $\alpha=5^\circ$, and
$\dot E_j\simeq 1.2\times 10^{44} \erg\s^{-1}$ (Con2,see Table \ref{tab_unif}).
 The arrows represent the velocity of the flow, according to the five 
velocity ranges given in section 3. }
\label{uniform_T0_1_E^44_fast}
\end{figure}

Figure \ref{uniform_T0_1_E^44_fast} shows the case with higher jet velocity of 
$v_j=23250 \kms$, $T=0.1\Myr$, $\dot E_j\simeq 1.2\times 10^{44} \erg\s^{-1}$ 
(Con2,see Table \ref{tab_unif}). As with wide jets (Paper I), fast jets of 
$v_j \ga 20,000 \kms$ (the exact speed limit depends on the other parameters) 
do not form fat bubbles, but rather the jets propagated close to the axis of 
symmetry (the x axis) in the manner of a narrow jet with an extensive cocoon.

\begin{figure}  
{\includegraphics[scale=0.29]{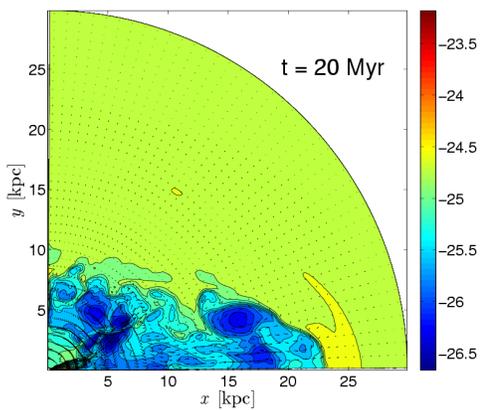}}
\caption{Density maps for a jet with a constant rate of changing precession 
angle, $5^\circ\leq\theta\leq65^\circ$, with a precession period of
$T_{\rm prec}=5\Myr$, half opening angle of $\alpha=5^\circ$,
$v_j=7750 \km\s^{-1}$, and $\dot E_j\simeq 1.6\times 10^{44} \erg\s^{-1}$ 
(Con7, see Table \ref{tab_unif}). The arrows represent the velocity of the 
flow as in Figure \ref{const_15deg_E^44}. The simulation was started with a 
precession angle of $\theta(0)=5^\circ$.}
\label{uniform_T5_E^44}
\end{figure}

In the case of constant rate of change in the precession angle, the jet spends 
equal time in small and large precession angles. At small precession angle the 
jet has more momentum per unit area, making it easier for the jet to break 
through the denser ICM. This can be seen in simulation with longer
precession periods, e.g., $T_{\rm prec} \ga 1\Myr$ (not shown here) where at
small precession angles the jet's propagation is easier than at larger angles,
resulting in a propagation along the symmetry axis, i.e., in the resemblance
of a narrow jet, as can be seen for example in Figure \ref{uniform_T5_E^44}
(Con7, see Table \ref{tab_unif}). No fat bubble is formed for a too long 
precession period. In all cases simulated in this section the typical 
temperature of the low density gas in the cavities or cocoon was of the 
order $10^8 \K \la T_b \la 10^9 \K$.

\section{RESULTS: RANDOM PRECESSION}
\label{sec:results3}

\begin{figure}
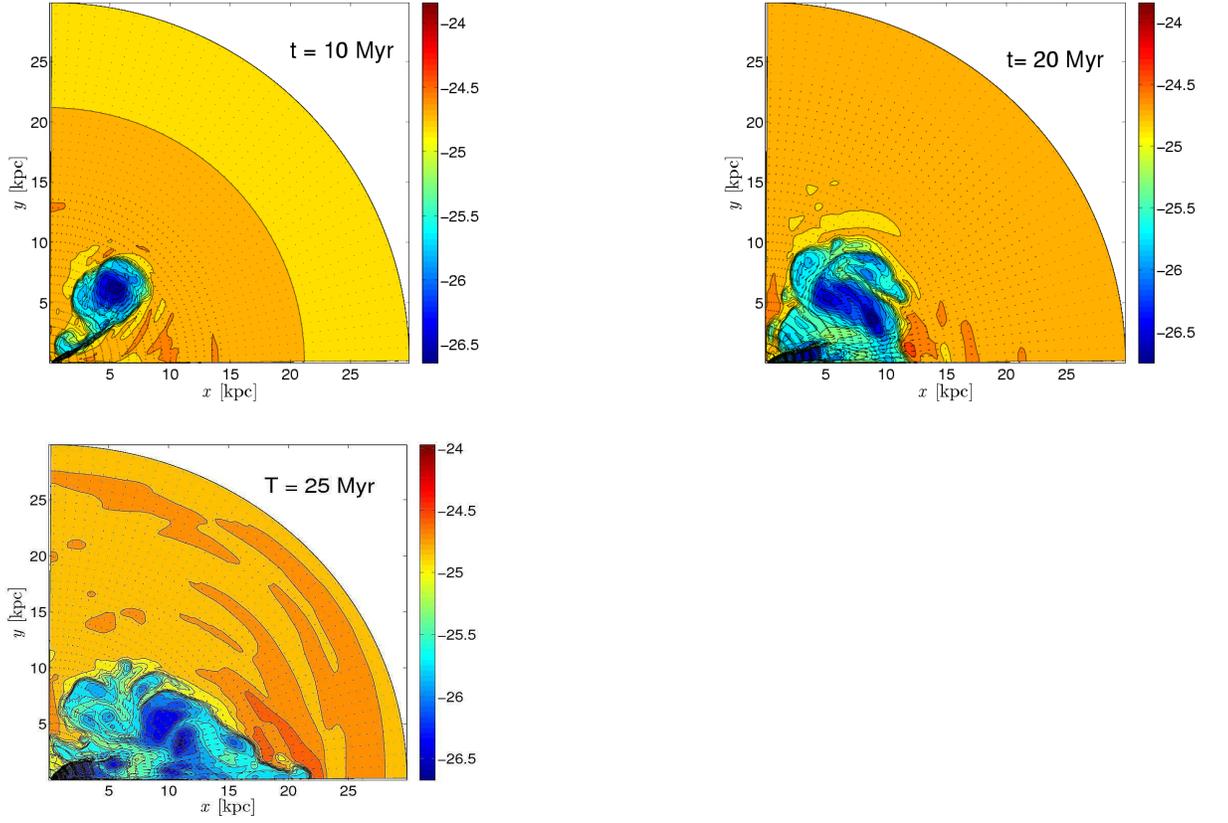
  
{\includegraphics[scale=0.29]{assafp2f10a.eps2}}
{\includegraphics[scale=0.29]{assafp2f10b.eps2}}
{\includegraphics[scale=0.27]{assafp2f10c.eps2}}
\caption{Density maps for a randomly precessing jet with
$T_{\rm prec}=30 \Myr$, $\alpha=5^\circ$, $v_j=7750 \km\s^{-1}$, and
$\dot E_j\simeq 1.8\times 10^{44} \erg\s^{-1}$ 
(Ran10, see Table \ref{tab_rand}). The jet's axis precess between
$\theta_{\rm min} = 5^\circ$ and $\theta_{\rm max} = 45^\circ$. The arrows
represent the velocity of the flow as in Figure \ref{const_15deg_E^44}.
The simulation was started with a precession angle of $\theta(0)=45^\circ$.}
\label{rand_T30}
\end{figure}

A more physically acceptable change in the precession angle is one in which
the jet covers a constant solid-angle per unit time, i.e., $d(\cos\theta)/dt$
is constant. Because the precession period about the symmetry axis is assumed 
to be very short, the jet spends longer periods of time at larger angels, in 
respect to the symmetry axis, than at small ones. The shorter time spent at 
small angels reduces the break-through period along the symmetry axis, as 
experienced in the uniform change in precession angel case (section 
\ref{sec:results2}). The reduction in the break-through period allows more 
matter to spread farther from the symmetry axis, in effect inflating a fat 
bubble of low density matter. This can be seen in Figure \ref{rand_T30} which 
shows the density map of a randomly precessing jet with a precession period of 
$T_{\rm prec}= 30 \Myr$ (Ran10, see Table \ref{tab_rand}).
The other parameters are $v_j=7750 \kms$,
$\dot E_j\simeq 1.8\times 10^{44} \erg\s^{-1}$, and
the boundary of the precession angle (of the jet's axis) are
$5^\circ< \theta <45^\circ$. The jet starts its precession off- axis and
inflates a toroidal cavity as seen at $t=10 \Myr$. At $t=20 \Myr$ we see
that the cavity is in the shape of a fat ellipsoid bubble. At $t=25 \Myr$ the
bubble is more or less spherical with a radius $R \sim 20 \kpc$. There is
a flow of low density matter to the equatorial plane, but it is minimal.

\begin{figure}
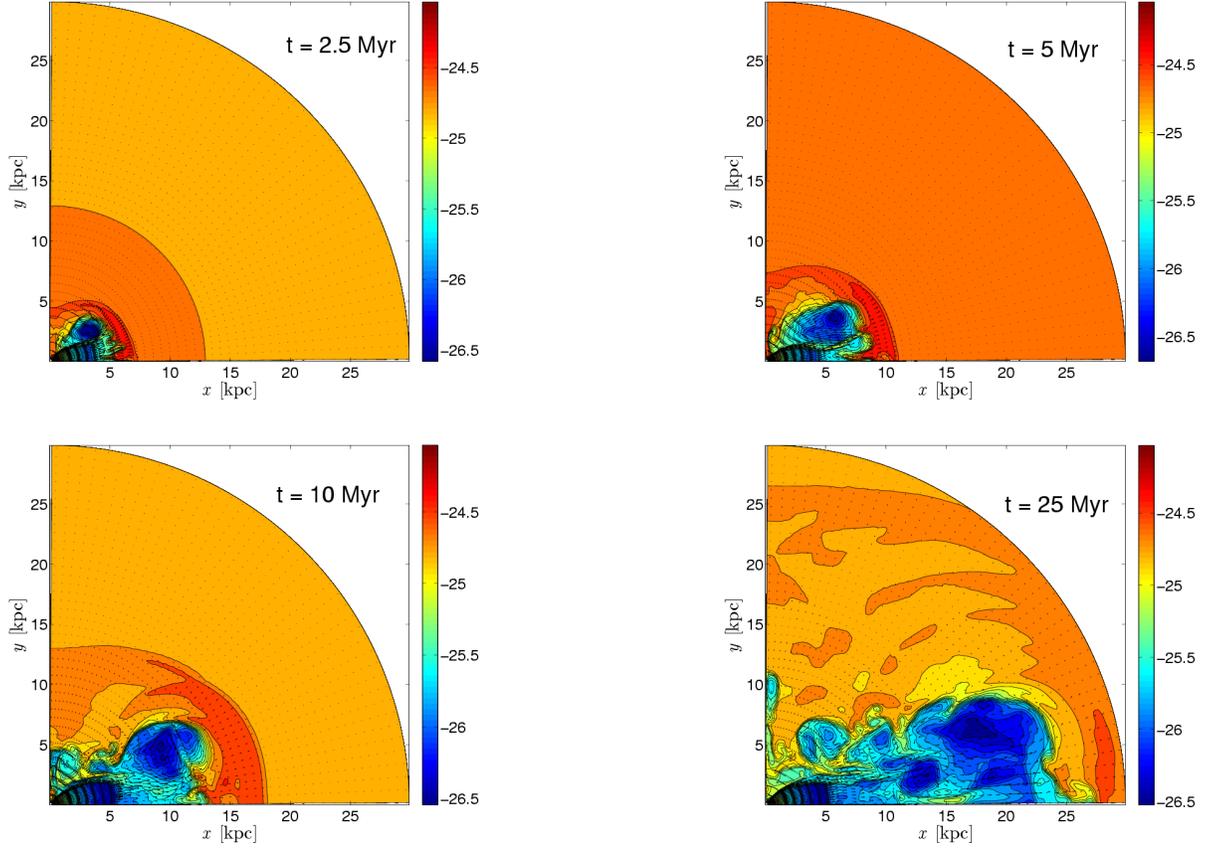
  
{\includegraphics[scale=0.29]{assafp2f11a.eps2}}  
{\includegraphics[scale=0.29]{assafp2f11b.eps2}}  
{\includegraphics[scale=0.29]{assafp2f11c.eps2}}
{\includegraphics[scale=0.29]{assafp2f11d.eps2}}
\caption{Density maps for a randomly precessing jet with precession period of
$T_{\rm prec}=0.1 \Myr$, jet's opening angle of $\alpha=5^\circ$,
jet's speed $v_j=7750 \km\s^{-1}$, and one jet power of
$\dot E_j\simeq 1.4\times 10^{44} \erg\s^{-1}$ 
(Ran1, see Table \ref{tab_rand}). The jet's interaction with
the ICM is similar to that of a wide jet with half opening angle of
$\alpha\sim 50^\circ$ (Paper I). The arrows represent the velocity of the
flow as in Figure
\ref{const_15deg_E^44}. }
\label{rand_T0_1}
\end{figure}

As in the constant rate of change case, the random precessing jet with a short
precessing period, $T_{\rm prec}=0.1 \Myr$ (Ran1, see Table \ref{tab_rand}) 
resulted in an interaction with the ICM similar to that of a wide jet with a 
half opening angle of $\alpha\sim 50^\circ$.
Figure \ref{rand_T0_1} shows the density maps for this case.
All along the simulation the cavity inflated by the jet is more or less
spherical, though at $t=25 \Myr$ it gets a little elongated. In contrast to
observations there is a flow of low density matter to the equatorial plane.

For higher velocity jets ($v_j \ga 2 \times 10^4 \kms$) the results (not
shown here) of the interaction between the jet and ICM are the propagation
of the jet close to the symmetry axis (i.e., a narrow jet with a cross
section radius of $1-2 \kpc$) with an extensive cocoon of low density matter
shed by the propagating jet in the form of vortices. The cross section radius,
of the jet and the cocoon, is typically $3-5 \kpc$.

Over all, we conclude that rapidly and randomly precessing massive slow jets
can inflate fat bubbles, similar to those inflated by wide jets (Paper I).
In all cases the typical temperature of the low density gas in the cavities
or cocoon was of the order $10^8 \K \la T_b \la 10^9 \K$.

\section{DISCUSSION AND SUMMARY} \label{sec:discussion}

We showed that precessing slow jets can inflate fat bubbles attached
to the center of the galaxy clusters.
By slow jets we refer to supersonic but highly non-relativistic jets.
In our axisymmetrical simulations the 3D problem can be simulated with
a 2D grid, and the simulations were of jets precessing rapidly
in the $\phi$ coordinate around the symmetry axis (the $\phi$ coordinate
is not included in our simulations).
Namely, the rotation time of the jet's axis around the symmetry axis
(the $x$ axis in our figures) is much shorter than the time
over which the precessing angle $\theta$ varies.
the length of our simulations was chosen to match the $10-50~$Myr age of most
observed bubbles (Birzan et al. 2004; McNamara \& Nulsen  2007).

The main criteria we find for the inflation of fat low density bubbles by
precessing jets are:
\begin{enumerate}
\item The jet's velocity should be $v_j \sim 10^4 \kms$.
Using our results obtaned here and our results from Paper I,
we conclude that the range over which slow jets can inflate fat bubbles is
$3000\kms \la v_j \la 2 \times 10^4 \kms$.
Jets with higher velocities form narrow expanding jets with extensive cocoons.
These jet velocities form bubbles with interior temperatures of
$ 10^8 \K \la T_b \la 10^{10} \K$.
\item For a jet with a fixed precession angle $\theta$ (measured
from the the symmetry axis; $x$ in our figures), this angle should not be
too small or too large.
A small angle causes the jet to propagate along the symmetry axis, and a large
angle leads to the bending of the jet towards the equatorial plane.
The constraint on the precessing angle is $30^\circ \la \theta \la 50^\circ$
\item The jet should spend more time at large angles in order to reduce
the break-through along the symmetry axis that happens when it's
precessing angle is small. This favors a case with random precession, i.e., 
the jet axis covers constant solid angle per unit time and the precessing 
angle is bound in the range 
$ 0^\circ \le \theta \le \theta_{\rm max} \simeq 50^\circ$.
\item For any prescribed precession behavior and parameters that can form
a fat bubble, the maximum precessing angle should be quite large,
$\theta_{max} \sim 30-70^\circ$.
\item The slow velocity and large energy of the jet inflating fat bubbles
require that the two opposite jets carry large amount of mass. The two jets
together can expel back to the ICM a mass at a rate of
$\dot M_{\rm back} \simeq 1-50 M_\odot \yr^{-1}$.
\end{enumerate}

Let us elaborate on these points, {{{and on the physics behind them.
As discussed in previous papers (Soker 2004; 2006; Paper I), the 
basic condition for a jet to inflate a fat bubble is that the jet's head 
will reside inside the bubble, or, if the jet's head is outside the bubble,
that the jet's head will not ''run away'' from the expanding bubble. For 
example, in the case of constant large precessing angle (Run 'Fix10' in Figure 
\ref{const_not_good}), the jet's head revolves around the symmetry axis but at 
a large distance. It inflates a local low density region, a torus around the 
axis. But this torus expands too slowly to inflate a fat bubble. Namely, the 
motion of the jet's head is too fast for the expanding shocked gas. This 
condition for the jet's direction not to escape from the expanding bubble 
formed by its shocked material is given in equation (17) of Soker (2006). 
This explains why the precessing angle cannot be too large (point 2 above).

The same principle holds for the jet's head not to expand too fast in the 
radial direction. If the expansion of the jet is concentrated within a small 
solid angle, e.g., a small precession angle, the large momentum flux (ram 
pressure) will result in a jet's head that moves radially faster than the 
expansion of the shocked gas. A long bubble will be formed instead of a fat 
bubble. This explains points 3 and 4 above and the lower limit on the 
precessing constant angle in point 2. Qualitatively, this condition on the 
solid angle of the expanding jet is given by equation (14) in Soker (2004) 
that was derived for a wide jet instead of a precessing jet, and was shown 
to hold in the numerical simulations of wide jets (Paper I).

Regarding point 1 above. If the jet's speed is too low, for a given energy 
it is too dense. As a result of that, the bubble expands too slowly and the 
jet's head moves radially too fast. This is also seen by equation (14) in 
Soker (2004), and was shown to hold for wide jets (Paper I). If the jet's 
material is too fast, the shocked jet's gas expands much faster than the 
jet's head. In that case the shocked low density gas fills the entire inner 
region, and the two opposite jets form a large elliptically-shaped bubble
instead of a bipolar (hourglass) structure of two fat bubbles. This is very 
similar to the case of wide angle (Paper I). Note that in some of the 
simulations no dense ICM gas is left in the equatorial plane near the 
center (e.g., Figure \ref{const_not_good}, see also Paper I). This shows that 
dense ICM in the equatorial plane near the center we find in some runs (e.g., 
Figures \ref{rand_T30}, see also Paper I) is real and not a numerical 
artifact.  }}}

{{{Point 5 has far reaching implications. }}} Are such a high mass outflow 
rates as we find here and in Paper I where we simulated wide jets, 
$\dot M_{\rm back} \simeq 1-50 M_\odot \yr^{-1}$,
compatible with observations? Are wide jets simulated in Paper I (or wide
precession angle simulated here) compatible with observations? The answer to
both questions seems to be positive. In Paper I we have already discussed
indications for AGNs that blow slow jets, some of them with wide angles
(Crenshaw \& Kraemer 2007; Behar et al. 2003; Kaspi \& Behar 2006). These
recent observational results (see also de Kool et al. 2001), and our numerical
results support the model where the feedback in both cooling flow clusters
and in the process of galaxy formation occurs mainly (but not solely, as
relativistic narrow jets also exist) by slow massive jets, as suggested and
discussed by Soker \& Pizzolato (2005). The massive jets imply that not only
energy, but mass as well is part of the feedback cycle (Soker \& Pizzolato
2005; Pizzolato \& Soker 2005).
Massive jets were also considered before by,
e.g., Begelman \& Celotti (2004) and Binney (2004), and were simulated by
Omma et al. (2004), who took the jet speed and mass outflow rate to be
$v_j = 10^4 \kms$ and $2M_\odot \yr^{-1}$, respectively. In the simulation
of Heinz et al. (2006) one jet has a mass loss rate of $35M_\odot \yr^{-1}$.
This implies that the two jets inject $70M_\odot \yr^{-1}$ into the ICM.
Without stating it, Heinz et al. (2006) followed the suggestion of Soker \&
Pizzolato (2005), that a large fraction or even most of the gas that cools
to low temperatures in cooling flow clusters gains energy directly from the
central black hole, and is injected back to the ICM.

The main effect discussed in the present paper is a dynamical effect with supersonic
velocities. Therefore, the temperature (and entropy) profile of the ICM has little
effect on the conclusions. In a forthcoming paper we will follow the bubbles as 
they buoy to larger distances ($\sim 100 \kpc$). There the entropy will have a 
crucial role, and a more realistic temperature profile will be used.

There are some discrepancies between our results and observations.
\begin{description}
\item[(1)] In many of the simulated cases, but not in all of them,
a non negligible low density matter flows to the equatorial plane and stays
there. This is generally not observed. It is quite possible that magnetic
tension will suppress this low density backflow. This is because the
backflow is expected to stretch magnetic field lines. In addition, gravity
might cause this low density equatorial gas to buoy outward. Gravity and
magnetic fields are not included in our simulations. We will include gravity
in future simulations. However, neither gravity nor magnetic fields are
expected to prevent the presence of hot low density gas in the equatorial
plane in cases of very fast jets where the backflow toward the equator is
strong.
\item[(2)] In some cases a torus shaped cavity is inflated at the beginning
of the simulation, achieving a spherical shaped cavity only at later times.
We remind the reader that we assumed a very short-period  precession about
the symmetry axis, in effect reducing the simulations dimensions to two. A
full three dimensional simulation of a randomly precessing jets (and even a
pulsed jet, whose activity is turned on and off), might result in the
inflation of low density small bubbles and not a torus as in the 2D
simulations. These small bubbles merge on a short time to form a larger bubble.
We leave the investigation of this process to a full 3D simulations in the
future.
\end{description} \par

Concerning these discrepancies and the observations discussed above,
we suggest that in most, but probably not in all, cases fat bubbles are
inflated by wide jets or rapidly and randomly precessing jets.


\acknowledgements
We thank John Blondin for his immense help with the numerical code.
We thank Ehud Behar and Nahum Arav for helpful discussions regarding
the use of wide massive slow jets.
This research was supported by the Asher Fund for Space Research at the
Technion.


\begin{references}

\reference{} Alouani Bibi, F., Binney, J., Blundell, K., \& Omma, H.
2007, astro-ph/0706.2949

\reference{} Basson, J. F. \& Alexander, P 2003, MNRAS, 339, 353

\reference{} Begelman, M. C. \& Celotti, A. 2004, MNRAS, 352, L45

\reference{} Behar E. et al. 2003, ApJ, 598, 232

\reference{} Binney, J. 2004, in The Riddle of Cooling Flows in Galaxies and
Clusters of Galaxies, Eds. T. Reiprich, J. Kempner, and N. Soker, published
electronically at http://www.astro.virginia.edu/coolflow/proc.php
(astroph/0310222)

\reference{} Binney, J., Alouani Bibi, F., \& Omma, H. 2007, MNRAS, 377, 142

\reference{} B{\^i}rzan, L., Rafferty, D.~A., McNamara, B.~R., Wise, M.~W.,
 \& Nulsen, P.~E.~J.\ 2004,  ApJ, 607, 800

\reference{} Blanton, E. L., Sarazin, C. L., \& McNamara, B. R.
   2003, ApJ, 585, 227

\reference{} Blanton, E. L., Sarazin, C. L., \& McNamara, B. R., \& Wise M. W.
   2001, ApJ, 558, L15

\reference{}  Blondin J.M., Kallman T.R., Fryxell B.A., Taam R.E. 1990, ApJ,
356, 591

\reference{} Br\"uggen, M., Heinz, S., Roediger, E., Ruszkowski, M., \&
Simionescu, A. 2007, arXiv0706.1869



\reference{} Crenshaw, D. M., \& Kraemer, S. B. 2007, ApJ, 659, 250

\reference{} de Kool, M., Arav, N., Becker, R. H., Gregg, M. D., White, R. L.,
 Laurent-Muehleisen, S. A., Price, T., \& Korista, K. T.  2001, ApJ, 548, 609


\reference{} Fabian, A. C., et al.\ 2000, MNRAS, 318, L65 

\reference{} Heinz, S., Br\"uggen, M., Young, A., \& Levesque, E. 2006, MNRAS,
373, L65

\reference{} Heinz, S. Choi, Y.-Y., Reynolds, C.~S., \& Begelman , M.~C. 2002,
ApJ, 569, L79

\reference{} Heinz, S., \& Churazov, E. 2005, ApJ, 634, L141

\reference{} Kaspi, S. \& Behar E. 2006, ApJ, 636, 674

\reference{} Loken, C.\, Roettiger, K.\, Burns, J.\ O.\, \&
  Norman, M.\ 1995, ApJ, 445, 80

\reference{} McNamara, B. R., \& Nulsen P.E.J. 2007, ARA\&A, 45, in press

\reference{} Morita, U., Ishisaki, Y., Yamasaki, N. Y., Ota, N., Kawano, N.,
Fukazawa, Y., \& Ohashi, T. 2006, PASJ, 58, 719

\reference{} Omma, H., Binney, J., Bryan, G., \& Slyz, A. 2004, MNRAS, 348,
1105

\reference{} Pizzolato, F., \& Soker, N. 2005, ApJ, 632, 821



\reference{} Reynolds, C. S., McKernan, B., Fabian, A. C., Stone, J. M., \&
Vernaleo, J. C. 2005, MNRAS, 357, 242

\reference{} Rodr\'iguez-Mart\'inez, M., Vel\'azquez, P. F., Binette, L., \&
Raga, A. C. 2006, A\&A, 448, 15

\reference{} Ruszkowski, M., Ensslin, T. A., Br\"uggen, M., Heinz, S., \&
Pfrommer, C. 2007, MNRAS, 378, 662



\reference{} Soker, N. 2004, A\&A, 414, 943  

\reference{} Soker, N. 2006, astro-ph/0608554

\reference{} Soker, N., \& Bisker, G. 2006, MNRAS, 369, 1115

\reference{} Soker, N., \& Pizzolato, F. 2005, ApJ, 622, 847

\reference{} Sternberg, A., Pizzolato, F., \& Soker, N. 2007, ApJ, 656, L5 
(Paper I)

\reference{} Stevens, I. R., Blondin, J. M., \& Pollock, A. M. T.
           1992, ApJ, 386, 265

\reference{} Vernaleo, J. C. \& Reynolds, C. S. 2006. ApJ, 645, 83

\reference{} Vrtilek, J. M., Grego, L., David, L. P., Ponman, T. J., Forman,
W., Jones, C., \& Harris, D. E.  2002, APS, APRB, 17107

\end{references}
\end{document}